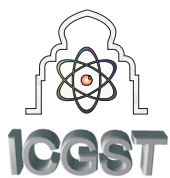

www.icgst.com

# Fostering of innovative usability testing to develop mobile application for mobile collaborative learning (MCL)

Khaled Elleithy, Abdul Razaque
*Wireless and Mobile communication Laboratory (WMC)*
*School of Engineering at the University of Bridgeport, Park Avenue, Bridgeport, CT, USA*
[elleithy,arazaque]@bridgeport.edu

## Abstract
Emergence of latest technologies has diverted the focus of people form Computer-Supported Collaborative Learning (CSCL) to mobile-supported collaborative learning. MCL is highly demanded in educational organizations to substantiate the pedagogical activities. Some of the MCL supportive architectures including applications are introduced in several fields to improve the activities of those organizations but they need more concise paradigm to support both types of collaboration: synchronous and asynchronous. This paper introduces the new pre-usability testing method that provides educational support and gives complete picture for developing new pedagogical group application for MCL. The feature of application includes asynchronous and synchronous collaboration support. To validate the features of application, we conduct the post usability testing and heuristic evaluation, which helps in collecting the empirical data to prove the effectiveness and suitability of Group application. Further, application aims to improve learning impact and educate people at anytime and anywhere.

## 1. Introduction
Today's people are raised with new emerging technologies as a phenomenal part of their routine lives. Mobile learning has greatly paid the attention of the new generation toward educational environment because mobiles are easily accessible, ubiquitous and portable devices and used by the majority of people anywhere and anytime. Mobile telephony connects the large numbers of potential and sophisticated learners through communication networks. This situation indicates that there is high potential for increasing the pedagogical learning with mobile devices [6]. MCL is a highly multidisciplinary learning paradigm around the world [1].

It attracts the people through new phenomenal way and its pedagogical, theoretical, technical and organizational structures being deployed [4] & [5]. MCL initiates a new rationales and paradigms to deliver learning materials into our daily life [8]. Many mobile communication frameworks promote the mobile learning through the learning portal by using Internet, sending SMS and voice communication. MCL could be made more effective due to convergence of promising interactive features of audio, video, web and new emerging mobile technologies in one package. Many pilot projects have already been launched to meet the requirements of distance education as part of e-learning to satisfy the demand of information and communication technology (ICT) for promoting the learning environments [7]. The main focus of the projects is to develop specific applications with limited features such as introduction of courses and quizzes.

The review of some usability testing and empirical survey on the latest status of MCL that explores alternatives to help the educational institutions to fulfill major functions of processing, storage and disseminating information that can be deployed with the issues of real life [9] & [10]. However existing literature is not sufficient for developing more powerful and general mobile application to meet most of the basic user requirements. The leading mobile application should have some interactive features to support self-directed, cognitive and constructivist quality for increasingly mobile learners to meet the educational targets. The thirst in exploring the new application, leads us to introduce new usability testing method with support of heuristic evaluation for devising the basic user requirement for MCL.

On the basis of collected information, we synthesize the new group application with very promising features to support pedagogical, organizational, social and other environments. Initially, validating the some of the important features of Group application conducted the post usability testing to find out the some core activities being used in MCL. The remaining part of this paper presents related work, pre-usability testing, which will help to explore the basic user requirements for designing and developing the new mobile application, section IV discusses the development of group application. initially to validate the some features of MCL, section V discusses the post usability testing and finally, conclusion of the paper.

## 2. Related work and Background study

There are many related contributions that provide strong support for importance of our new usability testing and group application. A major study done in Mobile Learning Environments (MLE) project was funded by Nordic Innovation Center. This project provided the forum to invite the people from businesses, universities and government organizations. The focus of the project was to introduce the pedagogical game-like practices in primary school systems of Nordic countries. The tests were conducted at Denmark, Finland and Sweden. The encouraging outcome of the project was to motivate the researchers to design pedagogical tools and applications [11].

The various types of mobile applications with general patterns of interactions are suggested in [12]. The paper also focuses on user study with low-fidelity and conduct paper-prototypes to analyze them. The results provide the guidelines and best practices for the design and development of general mobile applications [12]. The theme of the paper is to introduce an idea for development process of mobile application. The work is handy but it does not target any issues faced by users during the MCL. The Authors in [13] present the basic method of data collections such as interview, observation, questionnaire, survey and verbal protocol, which are employed for conducting the usability test of mobile application. The developed application covers the limited aspects for collaboration. Another contribution in [14] targets to desktop computing and subsequently leads for development of the conceptual models that helps for evaluation of mobile phone application. The authors mainly create the usability metric to analyze the mobile application.

Quality in Use Integrated Measurement (QUIM) is presented in [15]. QUIM helps the beginner who has trivial knowledge about the usability. This model covers 10 factors, which are categorized into 26 criteria. The model provides 127 metrics to measure the criteria. The model also helps in identifying the problem of working applications. However, the model is supporting the beginning process of usability testing but needs to be validated.

The authors in [16] have proposed a prototype for their participatory design project supported by inter generational design group to build mobile application and incorporate into iP Phone and iPod touch environment. The contribution provides the opportunities to bring the children and grandparents together by reading and editing the books. They focus on specific filed but work is not handling the issues required for effective MCL.

The work already done in previous papers do not provide the detailed information for building very promising application for MCL that may handle the issues for delivery of large rich multimedia contents (video-on-demand), administrator rights to teachers to check their courses and evaluate the progress of students, asynchronous and synchronous collaboration, support for multi model, provision for archive updating, user friendly interface, middleware support, virtual support.

## 3. Pre-usability testing & empirical results for design of MCL

The most influential and challenging tasks for designing and developing the mobile applications are to find out the basic needs of end users and how the requirements of users may be fulfilled and satisfied through applications. To meet the target of users, usability testing method with heuristic evaluation is the best practice to figure out the acceptable solutions. The suitability of applications is evaluated later, when the prototypes are tested and implemented, however usability testing makes the work of bridging the users with applications.

### 3.1. Setting objective

In order to conduct ranking, we analyze various features of MCL; some of them are already found in literature and arrange them according to feedback obtained from human interaction. These features provide the foundation to create new ranking method based on pre-usability testing method. This method involves the both genders including students, teachers, teaching Assistant and Administrators. Furthermore, our ranking method is based on two phases. In the first phase, we collect the user requirements through literature survey, interviews, coordination with experts of field to get valuable suggestions and comments. In the second phase, we sort out most influential basic requirements of users through questionnaire.

### 3.2. Organization of participants

The sources of collecting the information are Face book, vista survey and personal interactions. The process of coordination and conducting the interviews continued for four days. We summarize the findings and comparing them with existing literature survey. Finally, we combine all features by using intersection strategy. The participants of 15 universities became the part of this task. This task involves 106 users including 58 students, 23 teachers, 14 teaching assistants and 11 administrators from both genders. It helps in collecting the following user requirements given as follows:

- Interface should be easy to use
- Asynchronous Collaboration
- Synchronous Collaboration
- To support multimodal MCL
- To provide Archive updating
- Should be User friendly interface.
- To get a help from middle ware
- To give virtual support
- To provide application sharing facility and make easy text communication
- To provide admission functionality, if any user wants to participate in middle of the session.
- The administrators should be provided the opportunities to record the collaborative activities of students and teachers during the whole session or any specific period of time
- To provide the opportunities for interactive and shard white board
- Users may need short start time for collaboration
- Server should provide content adoption service
- The Students should have alternative choices in selecting any topic for discussion
- The Students should have access to check the comments given by teacher regarding their performance and grades
- The teacher should include critical notes for the performance of each student after completion of MCL session and provide the feedback to improve in future
- To provide audio and video communication only
- To provide connectivity management support
- To provide the support for session management
- To provide the checking facilities to instructor to check the group members
- To provide the freedom of thoughts to participating group members
- Server should give the message of information updating
- To provide the facility of translation of audio, video and text to other languages
- Client should give notification of his/her availability
- To provide the support for user role
- Portfolio should be created in order to store an information regarding the course
- To include group manager component
- The methods of communication should be direct or mediated
- The available digital materials should be integrated easily
- Instructor should dedicate time to monitor the progress of participating members
- To provide the support to handle the shared information
- To provide privacy and safety
- To provide the facility to contact and invite the participating for collaboration
- The communication should be based on broadcast with support of multicasting
- To make small participating group for collaboration
- To be flexible to collect and extract the data.
- To provide text, graphs, images, audio and video services to meet the requirements of related course of study.

- The teachers should have complete access to administer their courses and evaluate the progress of students

### 3.3. Conducting Evaluation and selecting basic requirements

All selected items are evaluated by applying second phase of ranking method based on questionnaire with support of 248 participants. These participants belong to education, business, social organizations and common people. Vista survey, face book, personal relationships helped to reach to those people to be part of ranking for finding the most basic user requirements. This activity helps to meet the pedagogical requirements through MCL.

We have used five-level Likert item in questionnaire, which helps to specify the level of agreement for each item. The Likert method covers the Strongly Agree= 1, Agree= 2, Neutral/No Opinion= 3, Disagree= 4 and strongly disagree= 5. Finally, we get empirical results, which are useful to understand the type of applications required for designing and developing for MCL. We sort out only the items, which have "strongly agree", "agree" and "Neutral/No Opinion" response. The items against any disagreement or strongly disagreement response are received that are not considered as basic requirements.

We introduce scaling values against each response such as strongly agree = 100 points, agree = 75 Points and Neutral/ No Opinion = 50 Points. The points for each user requirements are calculated with following formula.

**Mean Points for Each basic requirements= (Strongly Agree + Agree + No Opinion)/3**

The formula helps in calculating the values for each items, which are given against each basic requirements explained in table 1.

All of these testing procedures are helpful to understand the type of applications required for MCL. Based on the above results, we introduce innovative client-server based prototypes and mobile application" group" that meet the pedagogical targets of end users.

Table 1. Mean Value for each Basic requirement

| Description of Basic Requirement | Mean Value |
|---|---|
| Should be easy to use | 100% |
| Should be User friendly interface | 100% |
| To be flexible to collect and extract the data. | 99.698% |
| To provide text, graphs, images, audio and video services to meet the requirements of related course of study | 99.497% |
| The teachers should have complete access to administer their courses and evaluate the progress of students. | 98.995% |
| To support multimodal MCL | 98.09% |
| The administrators should be provided the opportunities to record the collaborative activities of students and teachers during the whole session or any specific period of time. | 97.59% |
| The teacher should include critical notes for the performance of each student after completion of MCL session and provide the feedback to improve in future. | 97.48% |
| The Students should have access to check the comments given by teacher regarding their performance and grades | 97.289% |
| To provide the checking facilities for instructor to check the group members | 90.00% |
| The methods of communication should be direct or mediated | 89.35% |
| To provide privacy and safety | 88.35% |
| To provide the facility to contact and invite the participating for collaboration | 83.83% |

## 4. Group Application

MCL is a newly emerging revolution for pedagogical requirements for all type of educational institutions. It allows the users to get computer-based information through mobile devices.MCL supports portability, connectivity, context awareness and social interaction [2]. Mobile is successful influential tool for collaboration, allowing the students to share and update the information for obtaining the targeted pedagogical activities. From one side, mobile creates the bridge of opportunities, and from other side, limitations make the hindrance for effective MCL [17]. The limitations, which highly degrade the performance of portable handheld devices, are small size of screen, mobility, low resolution, navigation issue, limited memory and bandwidth. The emerging technologies make this task easier and accessible for all.

The development of general mobile application that covers the many features for MCL can resolve the deep desire of students to make possible collaboration with mobile anytime and anywhere. Therefore our pre-usability testing gives the complete picture for developing the mobile application that will help the students to get the requested contents from server to fulfill the course requirements. We focus to develop application "group" with support of SDK and Visual basic and running on android operating system (OS) shown in figure 1 with other working functional components, which consists of control option and delivery option. The control option performs the functionalities of add new contact, edit contact, delete contact, existing collaborative group (C-G) and make new C-G. Delivery option covers the function of receive and send. When user requests for contents from server side, uses delivery option and saves with help of store component.

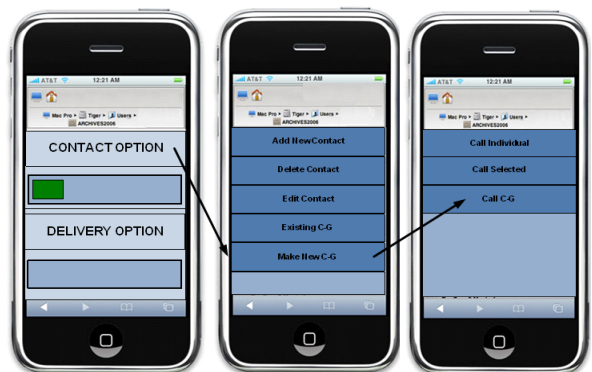

Figure 1: Group application with working functional components

The store component is core area of mobile where different type of data including file, audio and video are available to be obtained from server for MCL. The store component has also download option. The participants can use this option for downloading the data and storing onto Really Simple Syndication (RSS) 2.O for collaboration purposes. We developed the RSS (website), which provides the RSS feed to all the users to put and get contains frequently. Each section of store component manages the various files, audio and video. If once data is obtained that will be saved onto the store component of mobile and uploaded in relevant RSS feed. The upload section helps the user to store all types of data in RSS feed to be availed by participants member of same group for MCL. When user finishes the process of storing the data in its own RSS feed then informs the collaborative group by sending multicasting message to participate. The figure 2 shows main architecture that helps to get the contents and store contents.

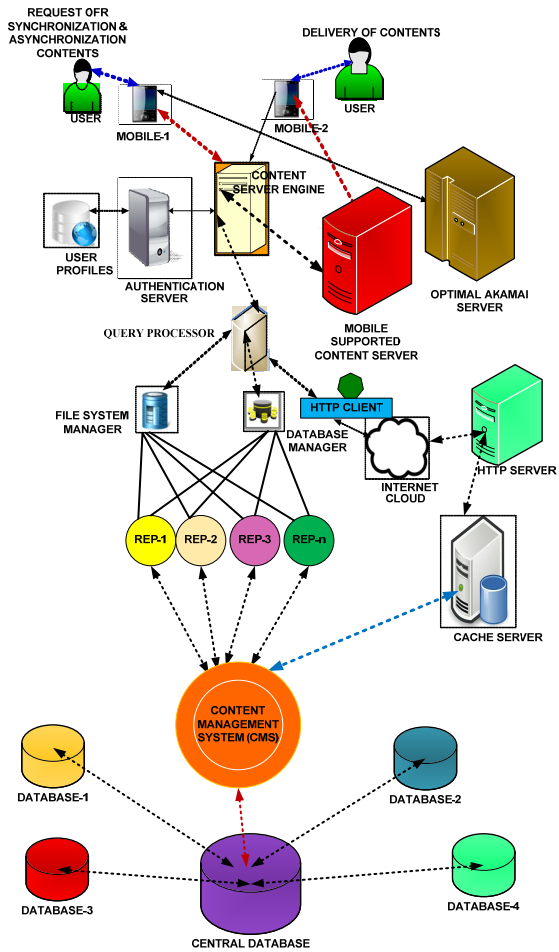

Figure 2: Architecture for MCL

When participants of group obtains the message for MCL and go to download the data from RSS feed of requested user. When they complete the process of downloading the required information from RSS feed of requested user, starts to process the MCL. If participating members of group need file for MCL, therefore open the file to obtain the information of the contents. In case the information of file is not sufficient to clear their understanding about the topic then play the video of related information.

The group application provides utility to manage and control the data with its relevant options. If type of data is based on only "text" then file option is used, otherwise audio and video options are used. When each of collaborative members get to know about topic, subsequently starts the process of MCL by using existing collaborative group (Existing CG) option shown in figure 3. The process can be supported by using H.323 at the server side to facilitate the real-time audio and video communication. The H.323 protocol will help to collaborative members to discuss the contents after reading and watching the video of related information. The important feature of H.323 is provision of point-to-multipoint, which creates the opportunities for all to participate simultaneously. The Quality of service can also be maintained by using H.323 [3]. With implementation of this application, MCL can be more successful with various promising features.

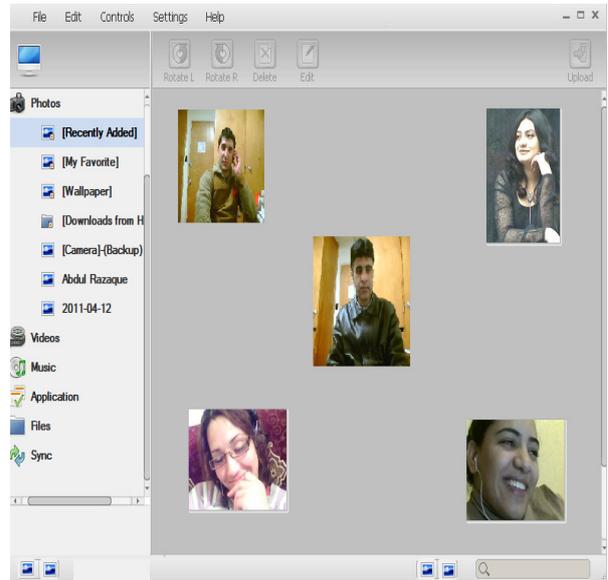

Figure 3: Participants of groups, sharing information through MCL

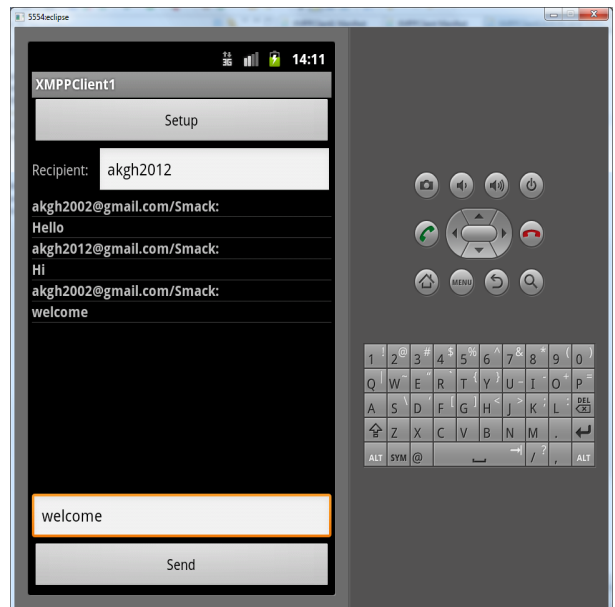

Figure 4: Recording the performance of various features of "Group" application

## 5. Post usability testing & empirical results for group application

In this section, we discuss post usability testing for group application based on heuristics and questionnaires. We choose the field test method that was conducted in the main lobby of North Hall in Bridgeport University and 08 participants took a part in application testing process. The tests evaluate the usability of group application. With this application, the users can transfer files between laptops and the mobiles.

The usability of group application testing procedure invites the multiple types of participants because many are familiar and belonging to mobile and wireless communication field and fewer possess less expertise in this field but know how to use mobile devices to make MCL.

Some users are using the chat service process to validate the performance of features of applications on the basis of activities shown in figure 4.

We devise the testing method using three steps. First, we introduce the testing procedure from design phase to conducting the test. Second, make all the related operations of group application and finally, we give the questionnaire to all the participants based on 5-level Likert method, so that participants give the answers after completing the appropriate tasks shown in figure 5.

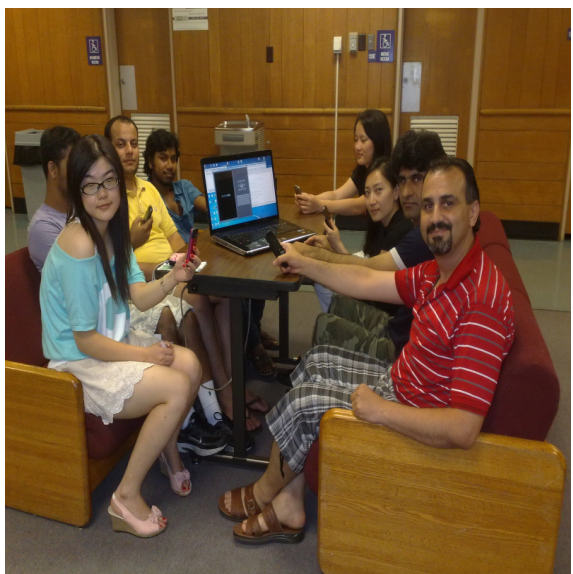

Figure 5: Testing the some features of group application for post usability testing

After completion of questionnaire, we arrange the meeting in which all the participants come and share their experience and give their impression regarding the test. We also get suggestions regarding the redesign and modification of application. When participants leave the room, we compile the results and arrange in form of table for better understanding shown in table 2.

## 6. Conclusion and future work

With emergence of innovative technologies in mobile communication and applications, the successful usability testing gains high importance in the design, development and deployment of interactive mobile applications. Therefore, it is essential to synthesize and adopt meaningful research methodologies in conducting the usability testing to analyze mobile applications. The contribution goes to creation of successful educational learning environment and development of highly supportive MCL application. This paper targets some of the major challenges being faced to enhance the student learning outcomes.

Table 2. Showing the comments of participants

| Post Usability Testing of fields | Comments of participants |
| --- | --- |
| Should be easy to use | Strong Positive=08<br>Positive=00<br>Neutral=00<br>Negative=00<br>Strong Negative=00 |
| Should be User friendly interface | Strong Positive=08<br>Positive=00<br>Neutral=00<br>Negative=00<br>Strong Negative=00 |
| To be flexible to collect and extract the data. | Strong Positive=06<br>Positive=01<br>Neutral=01<br>Negative=00<br>Strong Negative=00 |
| To provide text, graphs, images, audio and video services to meet the requirements of related course of study | Strong Positive=06<br>Positive=01<br>Neutral=01<br>Negative=00<br>Strong Negative=00 |
| To support multimodal MCL | Strong Positive=05<br>Positive=01<br>Neutral=02<br>Negative=00<br>Strong Negative=00 |
| The Students should have alternative choices for selecting any topic to discuss. | Strong Positive=03<br>Positive=02<br>Neutral=02<br>Negative=01<br>Strong Negative=00 |
| To provide the facility to contact and invite the members for collaboration | Strong Positive=02<br>Positive=03<br>Neutral=03<br>Negative=00<br>Strong Negative=00 |

The study of existing empirical literature and pre-usability testing based on interviews, questionnaire and heuristic evaluation highlight some features having importance for designing and developing the MCL application. Most of the mobile learning applications are limited to short-funded pilot projects, which were developed for special purposes or individual courses offered by their institutions. The most motivating and encouraging factor, which greatly impact on the success or failure of MCL depends on balancing of technology and human factors, how to create an effective pedagogical paradigm to support the various major basic user requirements for obtaining the contents through mobile devices. In addition, the way we spend our daily life in shopping, working, playing, learning highly being affected due to mobility of busy schedule. To control over some significant issues and providing an easy access of MCL at anytime and everywhere, introduce new group application.

It provides user friendly interface, access to administrators to record the collaborative activities of students and teachers, asynchronous and synchronous collaboration, multi model support, provision for archive updating, providing the facility to contact and invite the participants for collaboration, middleware support, virtual support and delivery of large rich multimedia contents (video-on-demand). Finally, we validate our claim by conducting the post usability testing and heuristic evaluation. In future, we will introduce more supporting features and software threads to complete MCL for different organizations.

## Biography

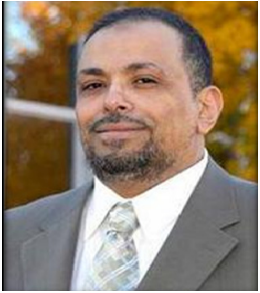

**Dr. Khaled Elleithy** is the Associate Dean for Graduate Studies in the School of Engineering and full professor in the dept: of computer Science & Engineering at the University of Bridgeport (UB). His research interests are in the areas of, network security, mobile wireless communications formal approaches for design and verification and Mobile collaborative learning. He has published more than one hundred twenty research papers in international journals and conferences in his areas of expertise.

Dr. Elleithy is the co-chair of International Joint Conferences on Computer, Information, and Systems Sciences, and Engineering (CISSE).CISSE is the first Engineering/Computing and Systems Research E-Conference in the world to be completely conducted online in real-time via the internet and was successfully running for four years. Dr. Elleithy is the editor or co-editor of 10 books published by Springer for advances on Innovations and Advanced Techniques in Systems, Computing Sciences and Software.

Dr. Elleithy received the B.Sc. degree in computer science and automatic control from Alexandria University in 1983, the MS Degree in computer networks from the same university in 1986, and the MS and Ph.D. degrees in computer science from The Center for Advanced Computer Studies in the University of Louisiana at Lafayette in 1988 and 1990, respectively. He received the award of "Distinguished Professor of the Year", University of Bridgeport, during the academic year 2006-2007.

.

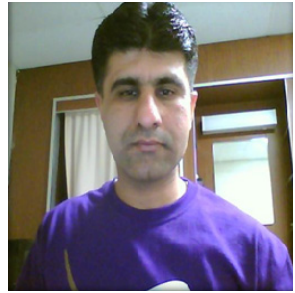

**Mr. Abdul Razaque** is PhD student of computer science and Engineering department in University of Bridgeport (UB). His current research interests include the design and development of learning environment to support pedagogical activities in open, large scale and heterogamous environments, collaborative discovery learning and the development of mobile applications to support mobile collaborative learning (MCL), the congestion mechanism of transmission of control protocol including various existing variants, delivery of multimedia applications. He has published over 40 research contributions in refereed conferences, international journals and books. He has also presented his work more than 10 countries. During the last two years he has been working as a program committee member in IEEE, IET, ICCAIE, ICOS, ISIEA and Mosharka International conference.

Abdul Razaque is member of the IEEE, ACM and Springer Abdul Razaque served as Assistant Professor at federal Directorate of Education, Islamabad. He completed his Bachelor and Master degree in computer science from university of Sindh in 2002. He obtained another Master degree with specialization of multimedia and communication (MC) from Mohammed Ali Jinnah University, Pakistan in 2008. Abdul Razaque has been directly involved in design and development of mobile applications to support learning environments to meet pedagogical needs of schools, colleges, universities and various organizations.